# Ground-state polariton condensation in 2D-GaAs semiconductor microcavities

M. Maragkou[1], A. J. D. Grundy[1], P. G. Lagoudakis[1*]

[1.] School of Physics and Astronomy, University of Southampton, Southampton SO17 1BJ, United Kingdom

We observe ground-state polariton condensation in a two dimensional GaAs/AlAs semiconductor microcavity under non resonant pulsed optical excitation. We resolve the formation of a polariton condensate by studying the spatial, angular, coherence, energy and transient dynamics of polariton photoluminescence. For high excitation densities we also observe a transition from the weak- to the strong-coupling regime in the time-domain and resolve the build-up of a coherent polariton state.

The possibility of high temperature Bose Einstein condensation (BEC) in the solid state offers an attractive alternative to ultracold atomic BEC. Polaritons in semiconductor microcavities, the admixture of an exciton and a cavity photon in the strong coupling regime are currently on the frontline of research in this field.[1] Due to their photon component, the de Broglie wavelength of polaritons is several orders of magnitude larger than that of atoms, allowing in principle for BEC even at room temperature.[2] However, unlike atoms, the polariton lifetime is limited by the photon cavity lifetime to a few picoseconds. Although the ultrashort lifetime prevents thermalisation with the host lattice, inter-particle interactions allow for rapid relaxation and formation of a macroscopically occupied ground state, usually referred to as polariton condensate.[3] Investigations in a CdTe microcavity demonstrated stark similarities of polariton condensation with non-equilibrium BEC.[4,5,6] Recently, an electrically pumped polariton diode was realised in planar GaAs microcavities paving the way to voltaic cell driven BEC in the solid state.[7,8,9] But despite the rich technology behind the development of high quality GaAs microcavities, polariton condensation in a 2D structure remains elusive even under optical pumping.[10]

In this Letter we report on the transient formation of ground-state polariton condensation in a 2D GaAs/AlAs semiconductor microcavity under non-resonant pulsed optical excitation. A threshold of the photoluminescence intensity with increasing excitation density is observed, while angularly resolved photoluminescence shows that polariton condensation occurs directly at the ground-state. Near field imaging of the polariton photoluminescence reveals a spatial collapse of the emission on the photonic disorder of the microcavity with the onset of polariton condensation occurring at polariton occupancy of approximately three. A transition from the weak- to the strong-coupling regime is observed at high excitation densities. The coherence of the emission in the weak coupling regime is significantly lower than in the strong coupling regime, allowing us to observe the build-up of a macroscopic coherent polariton state in the time domain.

The sample is grown by molecular beam epitaxy and consists of a λ/2 AlAs cavity of two $Al_{0.2}Ga_{0.8}As$/AlAs distributed Bragg reflectors with 16 (top) and 20 (bottom) pairs respectively. A set of four intra-cavity quantum wells is combined with two sets of four quantum wells inserted at the first antinode of the electromagnetic field in each Bragg mirror in order to increase the Rabi splitting at no cost to the cavity volume.[11] An unintentional wedge in the structure produces a variation of the exciton-cavity detuning. Here we work at a negative exciton-photon detuning of -4 meV. The sample is held in a wide-field view cold-finger cryostat and all experiments are performed at 7 K. In a previous study, under non-resonant continuous wave excitation photon lasing was observed in the same structure.[12] We attribute the transition to the weak coupling regime and the absence of any non-linear polariton effects in the strong coupling regime under CW excitation to sample heating. Here we eliminate such issues using 180 fs optical pulses at 730 nm focused down to a 50 um diameter spot and excite the sample at the first minima of the Bragg reflector.

Integrated photoluminescence intensity is recorded at normal incidence with ±1.5° collection angle and is spectrally resolved using a 1200 grooves/mm grating in a 55 cm spectrometer coupled to a cooled camera. The excitation density dependence reveals a non-linear increase of the photoluminescence intensity by $10^3$ at threshold [Fig. 1(a)]. The blue-shift of the emission energy with increasing excitation density is usually considered characteristic to the strong coupling regime and is interpreted as the result of repulsive polariton-polariton interactions.[13] Fig. 1(b) shows the energy shift of the peak emission as a function of the excitation density, which gradually flattens but remains below the bare cavity mode. The emission linewidth gradually broadens before threshold, it collapses at threshold and it broadens again above threshold with increasing excitation density as shown in Fig. 1(c). The rapid narrowing of the linewidth at threshold is attributed to the increased coherence in the polariton condensate.[14] Further broadening above threshold is partially due to increased scattering within the condensate. This observation is congruent to previous reports in CdTe and GaN microcavities.[15] However, as we show below, in the case of intermittent excitation time integrated photoluminescence measurements overestimate the linewidth of the emission due to the transient energy shift of the emissive states.

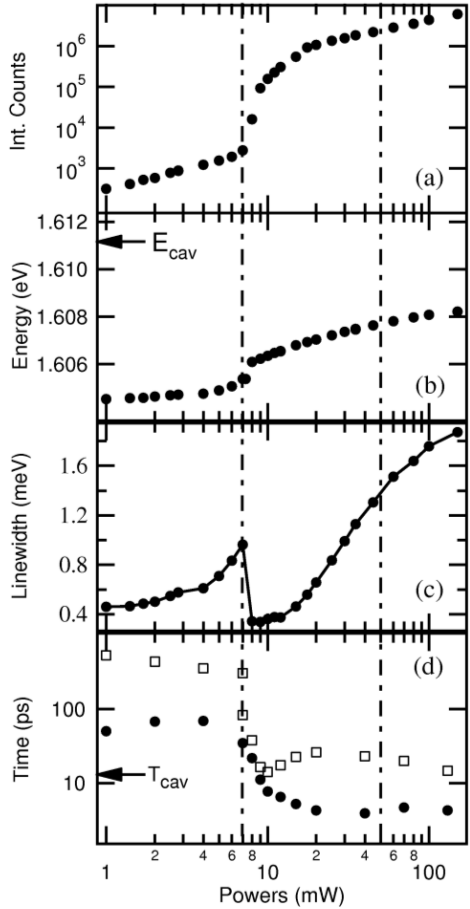

**FIG. 1** Power dependence of (a) the intensity, (b) the energy blueshift and (c) linewidth of the ground state emission as a function of the excitation power. (d) Rise (solid circles) and decay times (open squares) of the emission with increasing excitation power

The build-up and decay of the emission at normal incidence is resolved using a streak camera with an overall resolution of 3 ps. Fig.1(d) shows the rising time of the build-up of the emission (solid circles) and the corresponding decay time (open squares) as a function of excitation density. Below threshold the build-up (~90 ps) is driven by the relatively slow relaxation of polaritons at the ground-state (k=0) through exciton-phonon scattering and spontaneous exciton-exciton scattering that also dominate the photoluminescence decay dynamics.[16,17] Above threshold stimulated scattering accelerates the build-up of the emission and the decay dynamics are now limited by the cavity lifetime (~12 ps). Whereas the rise time follows a nearly monotonic dependence on the excitation density, we note that the decay time of the emission rapidly decreases above threshold nearing the cavity lifetime, while it increases for even higher excitation densities (open squares in Fig. 1(d)). We speculate that this is due to the depletion of the states in the exciton reservoir that feed the ground polariton state. At even higher excitation densities (to the right of the second vertical dashed line in Fig. 1(d)), the decay time re-approaches the bare cavity lifetime but as we show next, this trend is linked with the transition to the weak coupling regime.

Under pulsed excitation the transient dynamics of polaritons in microcavities result in time dependent energy shifts of the emission. We employ simultaneous energy- and time-resolved photoluminescence measurements using a 1200 grooves/mm grating in a 35 cm spectrometer coupled to a streak camera. Below threshold the emission energy corresponds to the ground polariton state shown in Fig. 2(a). The horizontal dashed lines depict the bare cavity and bare heavy hole exciton energies. The photoluminescence decay extends to 1.5 ns and is dominated by the relaxation of polaritons through exciton-exciton and exciton phonon scattering. It is possible that a residual electron population also contributes to relaxation via electron-exciton scattering.[18] The linewidth of the emission remains constant throughout the decay [Fig. 2(b)]. At threshold we can clearly resolve the transient dynamics of the emission energy that is initially blue-shifted while remaining well below the bare cavity mode and then gradually red-shifts with decreasing polariton population [Fig. 2(c)]. It is also evident that the linewidth of the emission is almost half the linewidth below threshold and that it remains virtually unchanged with time [Fig. 2(d)]. We note that the dynamic bandwidth of the streak camera does not allow us to resolve residual emission when the polariton dynamics return transiently to the linear regime.

From the transient dynamics of the emission energy and the linewidth in the non-linear regime it becomes apparent that integrated measurements of the photoluminescence under pulsed excitation produce only an upper estimate of the emission linewidth. This can also explain the broadening of the linewidth above threshold. Indeed at short times the dynamic blue-shift of the emission energy increases with increasing the excitation density. However, the emission energy of the polariton condensate always relaxes to the ground state of the polariton dispersion in the linear regime. This results in an increased time integrated linewidth with increasing excitation density when the instantaneous linewidth remains almost unchanged [Fig. 2(d)].

The most compelling evidence that the observed non-linearities in semiconductor microcavities occur in the strong coupling regime is produced when a clear transition to the weak coupling regime can be demonstrated for high excitation densities. In this study, for excitation densities six times above threshold, we observe an asymmetric spectral broadening of the emission developing on the high energy side of the polariton line. At excitation density 15 times above threshold this feature can clearly be resolved both spectrally and in the time domain as it precedes the emission from the ground polariton state [Fig. 2(e)]. The transient nature of this measurement allows us to resolve the evolution of the photoluminescence from the weak coupling regime, where the emission nears the bare cavity mode, directly into the non-linear strong coupling regime. Interestingly, in the weak coupling regime the emission linewidth is broader than the linewidth in the non-linear strong coupling regime. As the number of carriers in the quantum wells is reduced with time we observe a clear transition to the non-linear strong coupling regime. While the two regimes overlap the total emission linewidth is increased [Fig. 2(f)]. With the system evolving exclusively into the non-linear strong coupling regime the linewidth is narrowing to the value we observe above threshold but at lower excitation densities. This is the first observation of the transient build-up of a macroscopic coherent polariton state that pictures the formation of a polariton condensate spectrally and in the time domain.

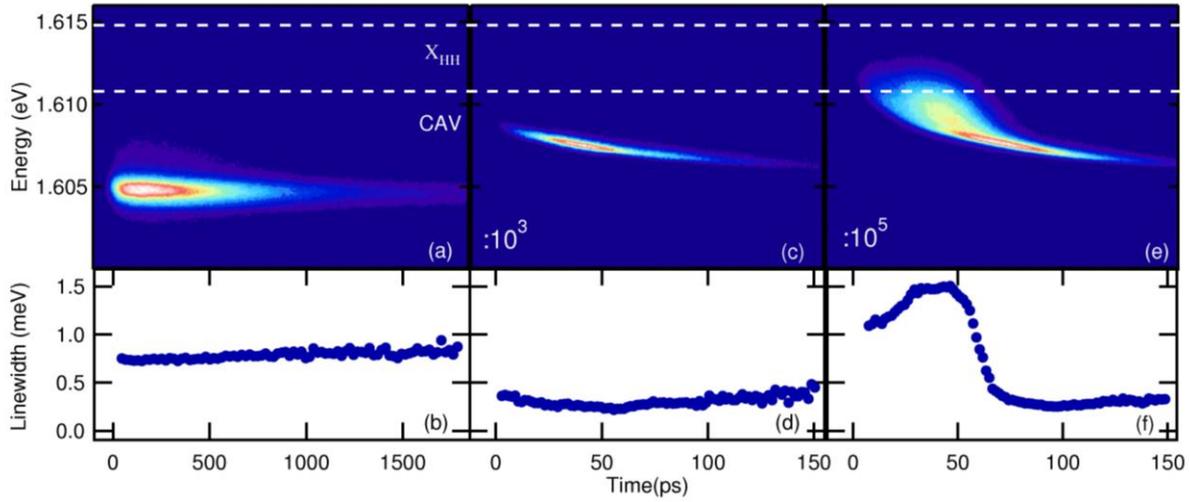

**FIG. 2** Time and spectrally resolved photoluminescence spectrum and the corresponding linewidth for excitations powers (a, b) below threshold, (c, d) at the non linear strong coupling regime and (e, f) at the weak coupling regime.

The inference of coherence from the linewidth of the emissive states is further endorsed by measuring directly the first order coherence of the photoluminescence both in the weak and strong coupling regime. For that purpose we build a Michelson interferometer and record interference in the far field.[19] Fig. 3(a,b) show the interference fringes at zero time delay in the weak and the non-linear strong coupling regime respectively. The corresponding contrasts are 10% and 40% exemplifying the higher coherence of the emission from the ground polariton state.[20] We also find that the decay time of the contrast of the observed fringes is longer in the ground polariton state (~8 ps) in comparison to the weak coupling regime (~4 ps) [Fig. 3(c)]. In the weak coupling regime the coherence lifetime is shown to be shorter than the cavity lifetime. In the non-linear strong coupling regime the measurement of the coherence lifetime is mostly limited by the rapid transient of the emission energy.

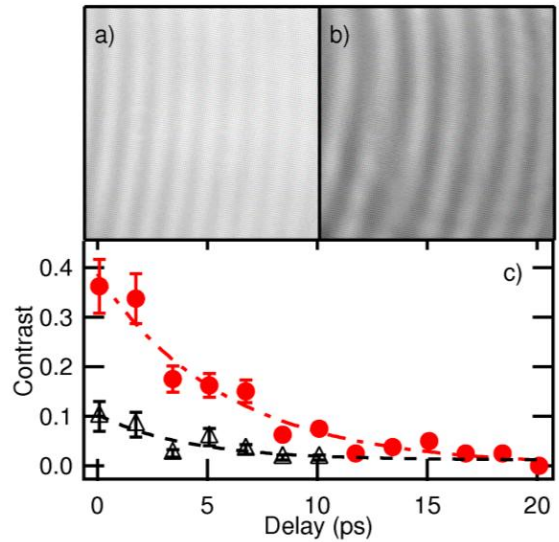

**FIG. 3** Far field interferograms of the ground state photoluminescence at zero time delay (a) in the weak coupling and (b) non-linear strong coupling regime. The dependence of interference contrast on time delay weak (black open triangles) and strong coupling (red solids circles) is shown in (c).

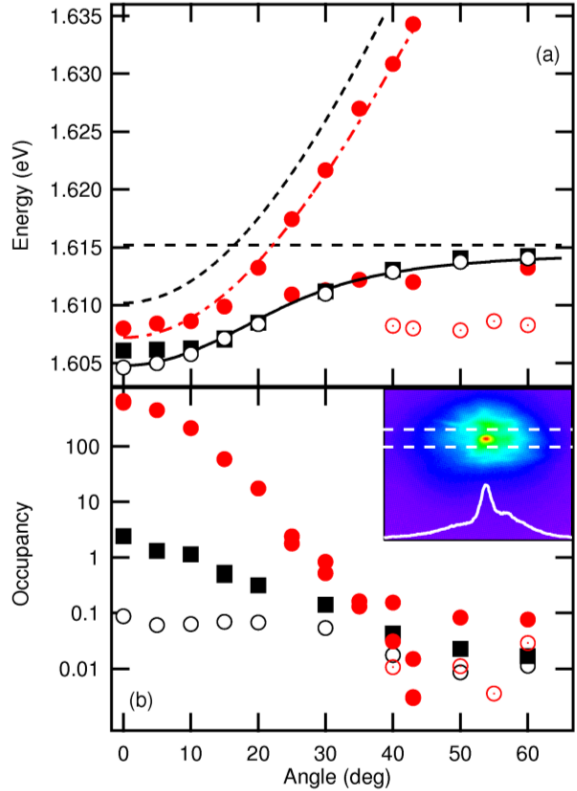

**FIG. 4** (a) Emission energy as a function of the detection angle for excitation powers (black open circles) below threshold, (black solid squares) at threshold and (red circles) at the weak coupling regime. The open red markers correspond to emission due to uncoupled excitons, observed at high angles. The black solid line correspond to the calculated lower polariton branch and the dashed black line to the heavy hole exciton and the cavity mode whereas the dashed red line to the cavity mode redshifted by 3meV. (b) Polariton occupancy as a function of the detection angle. (Inset) Real space image at threshold and the corresponding intensity profile.

The occurrence of non-linear photoluminescence intensity either at the ground-state or at a higher wavevector of the polariton dispersion varies in the literature and it appears to depend strongly on the excitation conditions.[21,22] Here we map the polariton dispersion below threshold, at threshold and in the weak coupling regime and correlate our findings with the spatial, near field pattern of the polariton emission. We build a goniometer that allows us to characterise the dispersion from normal incidence to 60 degrees and record the time integrated photoluminescence with 3 degrees collection angle using a spectrometer coupled to camera. The polariton dispersion in the linear regime (below threshold) is plotted in Fig. 4(a) (open black circles), where the solid line is the calculated lower polariton branch and the solid lines correspond to the uncoupled bare heavy hole exciton and cavity modes. At threshold we observe that the polariton dispersion blue-shifts to create a flat region between zero and 10 degrees, with the polariton states at higher angles being virtually unaffected [Fig. 4(a) (solid squares)]. To understand the angular broadening of the polariton dispersion near normal incidence we resolve the near field image of polariton photoluminescence at the excitation spot. Whereas the photoluminescence intensity below threshold follows the excitation intensity across the 40 μm spot, above threshold we observe a spatial collapse of the high intensity emission in a single spot of approximately ~3 μm diameter [inset in Fig. 4(b)]. A cross section of the excitation spot shows that most of the excited area remains in the linear regime that explains the observation of the linear polariton dispersion at wide angles. Also the spatial localisation of the polariton condensate in a 3 μm spot results in a symmetric angular broadening of ~19 degrees,[23] which can explain the observed flattening of the dispersion between zero and 10 degrees. From the intensity of the emission at threshold and the area of the emission we calculate the occupancy to be ~3.[24,25] We use this value to scale the occupancy for all other excitation densities. Figure 4(b) shows the occupancy of polaritons as a function of detection angle below threshold, at threshold and in the weak coupling regime. Evidently, polariton non-linearities spontaneously occur at the ground polariton state. In contrast to our observations in the strong coupling regime, in the weak coupling regime and for excitation densities 15 times above threshold the dispersion follows the bare cavity mode red-shifted by ~3 meV due to refractive index changes [solid red circles in Fig. 4(a)].[26] At high angles the bare cavity mode coexists with emission from other states red-shifted from the bare exciton mode [open red circles in Fig. 4(a)]. Emission from such states has been attributed to localised exciton states and their relative intensity is not monotonic with the in-plane wavevector [open red circles, Fig.4 (b)].[27]

In conclusion, we present compelling evidence of the formation of a macroscopic coherent polariton state in the time domain under non-resonant excitation. The transition between weak and strong coupling regime from a single excitation pulse allows for the first time the observation of the formation of ground-state polariton condensation in a planar microcavity.

The authors would like to acknowledge EPSRC and FP7 for funding.